# Design study on very low Beta spoke cavity for China-ADS[*]

LI Han(李菡)[1;2] SHA Peng(沙鹏)[2] DAI Jianping(戴建枰)[2;1)]
HUANG Hong(黄泓)[2] WANG Qunyao(王群要)[2]
1 (Graduate University of the Chinese Academy of Sciences, Beijing 100049, China)
2 (Institute of High Energy Physics, CAS, Beijing 100049, China)

**Abstract**: Very low Beta superconducting spoke cavity is one of the key challenges for China-ADS project. In this paper, a new structure of $3\beta\lambda/2$ spoke cavity is first presented. Its RF and mechanical properties are simulated using CST-MWS and ANSYS, and compared with the traditional $\beta\lambda/2$ spoke structure.

## 1. Introduction

Spoke cavity is a TEM-class and generally superconducting resonator. In addition to the inherent advantages of superconductors, the spoke cavity has more compact structure with the same frequency than the elliptical cavity, and higher shunt impedance than the half-wave resonator, etc.[1] Therefore, many large-scale accelerators, which need high power and high intensity protons, have been purposed to adopt spoke cavities, and the China-ADS linac is one of them.[2][3][4]

Figure 1 is the layout of China-ADS linac with the scheme of Injector-I.[4] The proton beam originates from an ECR ion source, and then accelerated to 3.2MeV by a CW normal conducting RFQ. From 3.2MeV to 178MeV, the beam is accelerated by three different types of superconducting spoke cavities (Spoke012, Spoke021, Spoke040). After that, the beam is accelerated by two types of superconducting elliptical cavities.

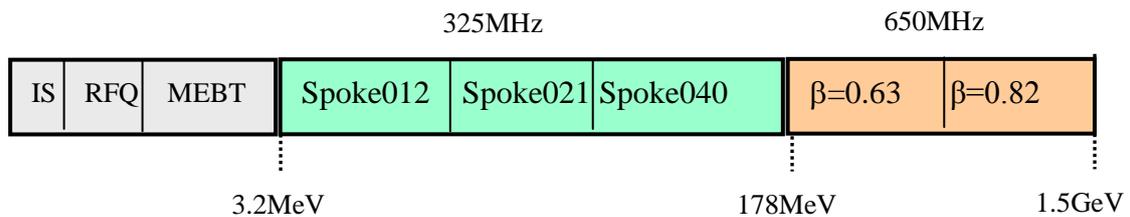

Figure 1: Layout of China-ADS linac with Injector-I

Among the three spoke cavities, Spoke012 is believed to be the most challenging one since it is of very low geometrical Beta, which is only 0.12. Lower Beta means thinner cavity and thus smaller areas of high magnetic fields, therefore presents naturally larger values of df/dP. In other words, the mechanical stability of low Beta spoke cavity against helium pressure fluctuations is inherently poor.[5]

In order to improve df/dP for a very low Beta spoke cavity, the typical approach is to develop sophisticated stiffening ribs, but it will not be discussed here. In this paper, we will investigate a new structure, whose length from gap-center to gap-center, $L_{g-g}$, is $3\beta\lambda/2$, instead of the traditional value of $\beta\lambda/2$, as illustrated in Figure 2. This new structure has different electric field pattern from the traditional one, as compared in Fig.3.

---

[*] Supported by the "Strategic Priority Research Program" of CAS, under Grant No. XDA03020600
1) E-mail: jpdai@ihep.ac.cn

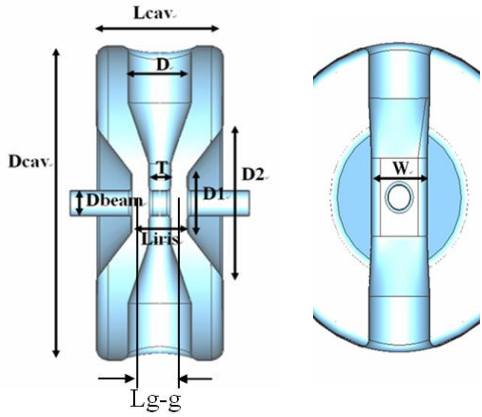

Figure 2: The cross section of a spoke cavity with the main geometric parameters: $L_{cav}$-cavity length, $L_{iris}$-iris to iris length, W-spoke width, T-spoke thickness, $D_{cav}$-cavity diameter, D-spoke diameter at base, $L_{g-g}$-length from gap-center to gap- center

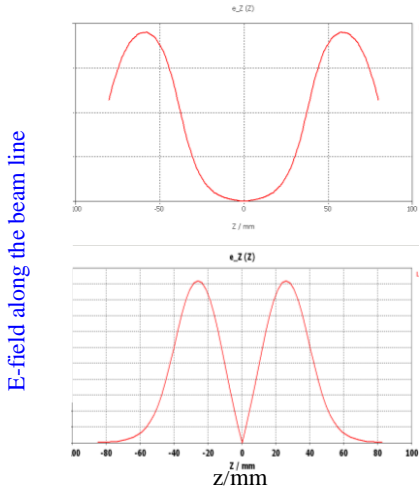

Figure 3: Different E pattern for traditional $\beta\lambda/2$ structure (bottom ) and $3\beta\lambda/2$ one （upper)

Led by the beam dynamics considerations, two different low Beta spoke cavities which may accelerate the 3.2MeV proton beam are chosen to be design-studied. One is the traditional Spoke012, and the other is the $3\beta\lambda/2$ spoke cavity, Spoke009T.

## 2. RF optimization of Spoke012 and Spoke009T

The RF design and optimization were done using Microwave Studio (MWS) software.[6]

For a spoke cavity, several parameters may typically be optimized, such as the spoke shape, spoke width-W, the gap ratio $T/L_{iris}$ and ratio $D/L_{cav}$. [7][8] As an example, Table 1 and Figure 4 show the optimization of the spoke shape for the traditional spoke012 cavity.

Table 1: RF parameters for different spoke shapes

| Spoke shape | Round | Elliptical | Racetrack |
|---|---|---|---|
| G ($\Omega$) | 51 | 53 | 70 |
| $R/Q_0$ ($\Omega$) | 124 | 135 | 93 |
| $E_{peak}/E_{acc}$[1] | 3.4 | 3.2 | 3.7 |
| $B_{peak}/E_{acc}$[1] (mT/(MV/m)) | 5.1 | 4.6 | 4.4 |

[1] $E_{acc}$ is the total accelerating voltage divided by $L_{iris}$

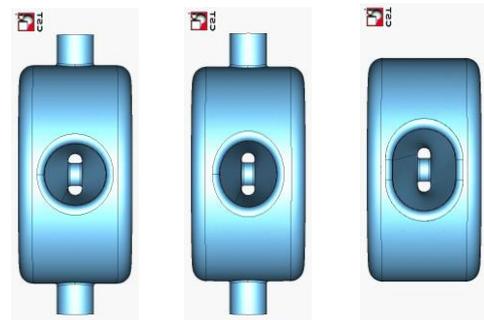

(a)Round   (b)Elliptical   (c)Racetrack

Figure 4: Three spoke shapes

Table 1 indicates that the elliptical spoke shape has better RF properties:

lower peak fields and higher R/Q. So we chose this shape for Spoke012.

For the new structure, Spoke009T, besides the above parameters, we have also investigated the influence of the effective cavity length ($L_{iris}$) while keeping $L_{g\text{-}g}$ constant, $L_{g\text{-}g}=3\beta\lambda/2=3\times0.087\times923/2=120$mm.

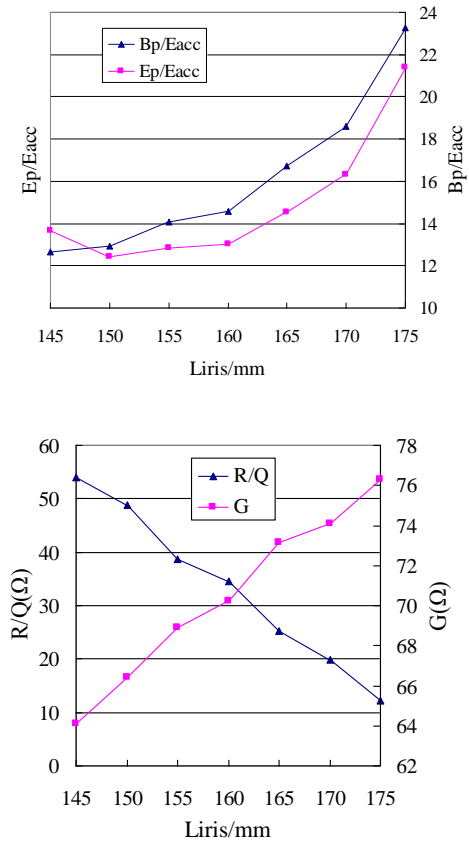

Figure 5: $B_p/E_{acc}$, $E_p/E_{acc}$, R/Q and G vs. $L_{iris}$ for Spoke009T

From Figure 5, $B_p/E_{acc}$, $E_p/E_{acc}$ and G increase greatly with $L_{iris}$ after 160mm, while R/Q decreases with $L_{iris}$. So 160mm is adopted for the $L_{iris}$ for a trade-off.

As a preliminary optimization result, the geometric parameters of Spoke012 and Spoke009T are confirmed and listed in Table 2.

Table 2: Geometric parameters of Spoke012 and Spoke009T

|  | $L_{cav}$ (mm) | $L_{iris}$ (mm) | T (mm) | W (mm) | $D_{cav}$ (mm) | $D_1$ (mm) | $D_2$ (mm) |
|---|---|---|---|---|---|---|---|
| Spoke012 | 170 | 74 | 29 | 89 | 438 | 81 | 219 |
| Spoke009T | 260 | 160 | 80 | 120 | 476 | 120 | 288 |

## 3. Comparisons between Spoke012 and Spoke009T

The comparison of RF properties is shown in Figure 6 and Table 3.

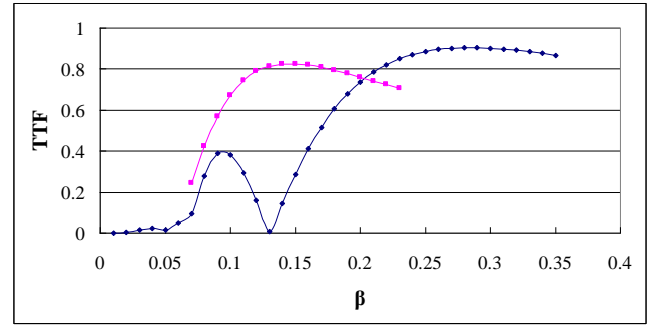

Figure 6: Transit time factor (TTF) at different β for Spoke012(rectangular) and Spoke009T(triangle)

Table 3: RF parameters of Spoke012 and Spoke009T

|  | f (MHz) | R/Q (Ω)@$\beta_g$ | $E_p/E_{acc}$ | $B_p/E_{acc}$ (mT/MV/m) | G (Ω) | TTF |
|---|---|---|---|---|---|---|
| Spoke012 | 325 | 135 | 3.2 | 4.6 | 53 | 0.77 |
| Spoke009T | 325 | 35 | 9.3 | 14.6 | 70 | 0.4 |

The mechanical properties of the two cavities without ribs have been simulated using ANSYS code[9], and the results are shown in Figure 7 and Table 4.

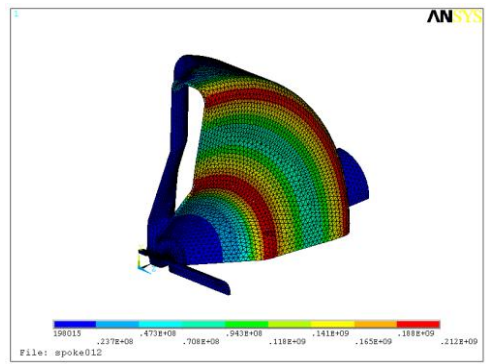

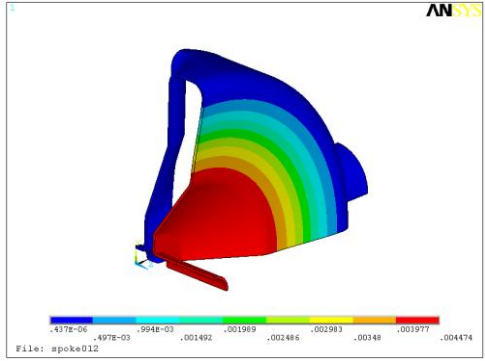

Figure 7 Simulation of stress (upper) and displacement (bottom) for Spoke012 at 1atm

Table 4: Mechanical parameters of Spoke012 and Spoke009T

|  | Spoke012 | Spoke009T |
|---|---|---|
| df/dP (kHz/Torr) | 18 | 6.8 |
| Freq. sensitivity for tuning (kHz/mm) | 1319 | 620 |
| Displacement sensitivity for tuning (mm/980N) | 0.58 | 0.44 |
| Peak displacement under 1atm with pipe free (mm) | 4.5 | 3.6 |

From the above analysis, it is found that compared with spoke012, spoke009T has thicker shell and about 3 times smaller df/dP, but lower TTF for the low energy proton beam whose β is about 0.1. And, as known, for the same accelerating voltage, lower transit time factor will induce higher peak field and larger cryogenic load.

## 4. Conclusion

Very low Beta spoke cavity is one of the key challenges for China-ADS linac, and it has no precedent in the world. Since the traditional structure for very low Beta spoke cavity has quite poor mechanical stability, a new 3βλ/2 structure has been studied. Preliminary design study shows that the new structure can significantly improve and enhance the mechanical properties of the cavity, but at the expense of poorer RF parameters. Next, we will optimize these two structures further and develop the Spoke012 prototype first.

## 5. Acknowledgement

The authors would like to thank Prof. TANG Jingyu and Dr. LI Zhihui of China-ADS physics group for the idea of 3βλ/2 structure and beneficial discussions.